# Relativistic extension of the complex scaling method

A. D. Alhaidari

*Shura Council, Riyadh 11212, Saudi Arabia*
AND
*Physics Department, King Fahd University of Petroleum & Minerals, Dhahran 31261, Saudi Arabia*
E-mail: haidari@mailaps.org

We construct a tridiagonal matrix representation for the three dimensions Dirac-Coulomb Hamiltonian that provides for a simple and straightforward relativistic extension of the complex scaling method. Besides the Coulomb interaction, additional vector, scalar, and pseudo-scalar coupling to short-range potentials could also be included in the same representation. Using that, we are able to obtain highly accurate values for the relativistic bound states and resonance energies. A simple program code is developed to perform the calculation for a given charge, angular momentum and potential configuration. The resonance structure in the complex relativistic energy plane is also shown graphically. Illustrative examples are given and we verify that in the nonrelativistic limit one obtains known results. As an additional advantage of this tridiagonal representation, we use it to obtain a highly accurate evaluation of the relativistic bound states energies for the Woods-Saxon potential (as a model of nuclear interaction) with the nucleus treated as solid sphere of uniform charge distribution.

## I. INTRODUCTION

Studying the properties of the resolvent operator (Green's function) associated with the scattering of a projectile by a target is essential to the understanding of both the structure of the target and the interaction of the projectile-target system. For example, bound states and resonance energies are identified with the poles of the Green's function $G(z) = (H-z)^{-1}$ in the complex $z$-plane, where $H$ is a "complexified" version of the Hamiltonian of the system. Dismissing subtle differences in perturbation theory between resonances and eigenvalues for degenerate states [1], it becomes obvious that the poles of $G(z)$ are the complex eigenvalues of $H$ in the $z$-plane. Resonance states are bound-like states that are unstable and decay with a rate that increases with the value of the imaginary part of the resonance energy.

In nonrelativistic quantum mechanics, the dynamical behavior of the state of the system in time is contained in the exponential factor $e^{-iEt}$, where $E$ is the nonrelativistic energy. For stable states, like the bound states, $E$ is real. However, for the decaying resonance states, $E$ is complex with negative imaginary part. Systems with hermitian Hamiltonians have no states with positive imaginary part for $E$, which would then blow up in time. Therefore, for systems with a self-adjoint Hamiltonian, energy resonances are located in the lower half of the complex energy plane. Sharp or "shallow" resonances are located below and close to the real energy axis in the complex $E$-plane. These are more stable. They decay slowly and are easier to obtain than broad or "deep" resonances that are located below, but far from, the real energy axis [2]. Most of the algebraic methods used for the study of resonances are applied directly in the complex energy plane, whereas most of the analytic investigations are done in the complex angular momentum plane [3]. The energy spectrum is the set of poles of the Green's function in the complex energy plane, which consists generally, for one-particle (single channel) Hamiltonian, of three parts:
1) Discrete set of real points on the negative energy axis corresponding to the bound states.
2) Discontinuity of the Green's function along the real positive energy line (a branch cut), which corresponds to the continuum scattering states[+].
3) Discrete set of points in the lower half of the complex energy plane corresponding to the resonance states.

The basic underlying principle in the various numerical methods used in the study of resonances is that the position of a resonance is stable against variation in all unphysical computational parameters.

Consequently, resonance energies are the subset of the poles of the Green's function $G(E)$, which are located in the lower half of the complex energy plane. One way to uncover these resonances, which are "hidden" below the real line in the $E$-plane, is to use the complex scaling (a.k.a. complex rotation) method [4]. This method exposes the resonance poles and makes their study easier and manipulation simpler. It has been used very successfully in nonrelativistic atomic and nuclear calculations of resonances. In this method, the radial coordinate gets transformed as $r \to re^{i\theta}$, where $\theta$ is a real angular parameter. The effect of this transformation on the pole structure of $G^\theta(E) \equiv \left(H^\theta - E\right)^{-1}$ in the complex $E$-plane, where $H^\theta$ is the complex-scaled Hamiltonian, consists of the following:
1) The discrete bound state spectrum that lies on the negative energy axis remains unchanged.
2) The branch cut (discontinuity) along the real positive energy axis rotates clockwise by the angle $2\theta$.
3) Resonances in the lower half of the complex energy plane located in the sector bound by the new rotated cut line and the positive energy axis get exposed and become isolated.

Figure 1 is a graphical representation of this process. However, due to the finite size of the basis set used in the calculation, the matrix representation of the Hamiltonian is finite resulting in a discrete set of eigenvalues (poles of the Green's function). Consequently, the rotated cut line gets replaced by a string of interleaved poles and zeros of the finite Green's function that tries to mimic the cut structure. Additionally, in this finite approximation the $2\theta$ cut line becomes deformed in the neighborhood of resonances due to the effect of localization in the finite $L^2$ bases in regions that are near to the resonance energies. Figure 2 is a reproduction of Fig. 1 but with a finite basis set. Now, the subset of the eigenvalues that corresponds to the bound states and resonance spectra remain stable against variations in all computational parameters (including $\theta$, as long as these poles are far enough from the cut "line"). For multi-channel scattering, on the other hand, there are as many cut lines (branch cuts) as there are channels. Complex scaling causes each cut line

---

[+] Generally, this part of the spectrum consists of a set of disconnected energy bands with forbidden energy gaps in between. Each band consists of continuous scattering states with energies bounded within that band.

to rotate about the corresponding channel's threshold energy point with an angle equals to twice the scaling angle of that channel [4].

The basis for the generalization of the complex scaling method to the relativistic problem was first outlined by Weder more than 30 years ago [5]. The mathematical details of this generalization were given later by Šeba [6]. Nonetheless, the implementation of the method on the relativistic problem has been largely ignored in the physics literature for a long time. We are aware of only two recent applications of the method. One is by Ivanov and Ho for computing resonances of Hydrogen-like ions in the presence of a uniform electric field [7]. The other is by Pestka *et al.* for obtaining the bound states energies of two-electron atoms within a variational Hylleraas-CI approach [8]. In this work, we present a general and systematic development of an algebraic extension of the complex scaling method to the relativistic problem. The Hamiltonian of the system will be taken to be the three dimensional Dirac-Coulomb Hamiltonian with an additional coupling to a finite range potential matrix. We assume spherical symmetry and consider three different types of coupling of the Dirac particle to the scattering potential. These are the vector, scalar, and pseudo scalar couplings. In the following section, we construct the spinor basis that results in a tridiagonal matrix representation for the reference Dirac-Coulomb Hamiltonian. The matrix elements of the scattering potential will be calculated in a finite subset of the basis using Gauss quadrature approximation [9]. In Sec. III, we show that this representation makes the application of the complex scaling method to the relativistic problem very simple and straightforward. The resonance structure in the relativistic complex energy plane will be shown graphically. Some potential examples are given and we show that in the non-relativistic limit we obtain known results. Moreover, we calculate the relativistic bound states energies for the Woods-Saxon potential (as a model for nuclear interaction) in the presence of the Coulomb interaction for a given set of physical parameters and for three different kinds of coupling: vector, scalar, and pseudo-scalar. In Sec. IV, we discuss our results and give some ideas about further developments to improve the accuracy of the method. A simple program code (RCS-07.1) was developed, using Mathcad®, to implement the relativistic extension of the complex scaling method and produce all results given in this work. A copy of the code is available upon request from the author.

## II. THE TRIDIAGONAL SPINOR REPRESENTATION

In this work, we consider the three dimensional relativistic scattering problem of spin $\frac{1}{2}$ charged particle (of mass $m$) with a massive target. The projectile-target system is described by the time-independent Hamiltonian

$$\mathcal{H} = \mathcal{H}_0 + \mathcal{V} \qquad (2.1)$$

The reference Hamiltonian $\mathcal{H}_0$ is taken to be the three dimensional Dirac Hamiltonian that may include coupling to an exactly solvable 4×4 potential matrix $\mathcal{V}_0$. It is permissible for this "reference potential" to have long-range interaction (e.g., the Coulomb potential). However, the scattering potential matrix $\mathcal{V}$ has finite range such that it is well represented by its matrix elements in a finite square integrable spinor basis. In the units $\hbar = m = e = 1$, the Compton wavelength is $\lambdabar = \hbar/mc = 1/c$, which could be written for an electron

---

® Mathcad is a software package developed by Mathsoft for general-purpose mathematical computations.

projectile as $\lambdabar = \alpha a_0$, where $\alpha$ is the fine structure constant and $a_0$ is the Bohr radius. In these units, the reference Dirac Hamiltonian for the charged spinor with coupling to the electromagnetic 4-vector potential $(A_0, c\vec{A})$ reads as follows [10]

$$\mathcal{H}_0 = \begin{pmatrix} +1 + \lambdabar^2 A_0 & -\lambdabar \vec{\sigma} \cdot (i\vec{\nabla} + \vec{A}) \\ -\lambdabar \vec{\sigma} \cdot (i\vec{\nabla} + \vec{A}) & -1 + \lambdabar^2 A_0 \end{pmatrix} \tag{2.2}$$

where the Hamiltonian is written in units of $mc^2 = \lambdabar^{-2}$ and $\{\vec{\sigma}\}$ are the three 2×2 hermitian Pauli spin matrices:

$$\sigma_1 = \begin{pmatrix} 0 & 1 \\ 1 & 0 \end{pmatrix}, \quad \sigma_2 = \begin{pmatrix} 0 & -i \\ i & 0 \end{pmatrix}, \quad \sigma_3 = \begin{pmatrix} 1 & 0 \\ 0 & -1 \end{pmatrix}. \tag{2.3}$$

Our choice of atomic units ($\hbar = m = 1$) over the conventional relativistic units (where $\hbar = c = 1$) is made to allow us to take the nonrelativistic limit, $c \to \infty$ (i.e., $\lambdabar \to 0$), in a very simple, intuitive, and straight-forward manner which is not possible in the latter units since $c = 1$. Additionally, it is easier to compare our results with those in atomic physics since the same system of units are used. Moreover, $mc^2 \to \infty$ is not a good measure of the nonrelativistic limit for position-dependent mass systems, which is an interesting problem that is becoming the core of an active field of research. This is because this limit could be violated in regions where the local mass distribution is infinitesimal despite the fact that the system is certainly nonrelativistic.

For the Coulomb interaction, where $A_0 = Z/r$ and $\vec{A} = 0$, and for spherically symmetric potential $\mathcal{V}(r)$, the angular components separate and the radial reference Dirac Hamiltonian becomes

$$\mathcal{H}_0 = \begin{pmatrix} +1 + \lambdabar^2 \frac{Z}{r} & \lambdabar\left(\frac{\kappa}{r} - \frac{d}{dr}\right) \\ \lambdabar\left(\frac{\kappa}{r} + \frac{d}{dr}\right) & -1 + \lambdabar^2 \frac{Z}{r} \end{pmatrix} \tag{2.4}$$

where $Z$ is the dimensionless electric charge coupling and where length is measured in units of $4\pi\epsilon_0 \hbar^2 / me^2$ (for an electron this unit is $a_0$). The spin-orbit quantum number $\kappa = \pm 1, \pm 2, ..$ and it is related to the orbital angular momentum quantum number $\ell$ by $\kappa = \pm\left(\ell + \frac{1}{2}\right) - \frac{1}{2}$. On the other hand, the radial scattering potential matrix is

$$\mathcal{V}(r) = \lambdabar \begin{pmatrix} \lambdabar V_+(r) & W(r) \\ W(r) & \lambdabar V_-(r) \end{pmatrix} \tag{2.5}$$

where $V_\pm = V \pm S$. $V(r)$ is the vector potential, $S(r)$ is the scalar potential and $W(r)$ is the pseudo-scalar potential. The reference Dirac-Coulomb problem described by $\mathcal{H}_0$ is exactly solvable. One such solution is obtained (for all energies) as an infinite sum of square integrable functions with expansion coefficients that are orthogonal polynomials in the energy [11]. However, the spinor basis in that solution is energy dependent; a property which is not desirable from numerical point of view. This is because any calculation in such a basis has to be repeated for all energies in the range of interest. Nonetheless, we will use that solution only as a guide to the construction of the spinor basis for the solution space of the present problem. We start by transforming the total radial Hamiltonian using the following 2×2 unitary matrix

$$\mathcal{U}(\varphi) = \exp\left(\tfrac{i}{2} \varphi \sigma_2\right) \tag{2.6}$$

where $\varphi$ is a real angular parameter such that $\sin\varphi = \pm\lambdabar Z/\kappa$ and where $-\frac{\pi}{2} \leq \varphi \leq +\frac{\pi}{2}$ depending on the signs of $Z$ and $\kappa$. The plus/minus sign in $\sin\varphi$ corresponds to the positive/negative energy solutions. In what follows, we consider only the positive energy solutions, where $\sin\varphi = \lambdabar Z/\kappa$. One can easily show that the negative energy solutions are obtained from the positive energy solutions by exchanging the upper radial spinor component with the lower and applying the following ($\mathcal{CP}$-) map

$$\kappa \to -\kappa,\ Z \to -Z,\ \text{and}\ W(r) \to -W(r) \tag{2.7}$$

If we write the transformed Hamiltonian as $H = \mathcal{U}\mathcal{H}\mathcal{U}^{-1}$, then the wave equation reads

$$(H - \varepsilon)\chi(r,\varepsilon) = 0 \tag{2.8}$$

where $\varepsilon$ is the relativistic energy in units of $mc^2 = \lambdabar^{-2}$. For bound states $|\varepsilon| < 1$, whereas for scattering $|\varepsilon| > 1$. The transformed reference Hamiltonian becomes

$$H_0 = \mathcal{U}\mathcal{H}_0\mathcal{U}^{-1} = \begin{pmatrix} \frac{\gamma}{\kappa} + 2\lambdabar^2\frac{Z}{r} & \lambdabar\left(-\frac{Z}{\kappa} + \frac{\gamma}{r} - \frac{d}{dr}\right) \\ \lambdabar\left(-\frac{Z}{\kappa} + \frac{\gamma}{r} + \frac{d}{dr}\right) & -\frac{\gamma}{\kappa} \end{pmatrix} \tag{2.9}$$

where $\gamma = \kappa\cos\varphi = \kappa\sqrt{1 - (\lambdabar Z/\kappa)^2}$. On the other hand, we write the transformed scattering potential as

$$U = \mathcal{U}\mathcal{V}\mathcal{U}^{-1} = \lambdabar\begin{pmatrix} \lambdabar U_+ & U_0 \\ U_0 & \lambdabar U_- \end{pmatrix} \tag{2.10}$$

where

$$U_\pm = \tfrac{1}{2}(V_+ + V_-) \pm \tfrac{\gamma}{\kappa}\tfrac{1}{2}(V_+ - V_-) \pm \tfrac{Z}{\kappa}W \tag{2.11a}$$

$$U_0 = \tfrac{\gamma}{\kappa}W - \lambdabar^2\tfrac{Z}{\kappa}\tfrac{1}{2}(V_+ - V_-) \tag{2.11b}$$

Now, we expand the solution of the wave equation (2.8) as $\chi(r,\varepsilon) = \sum_n C_n(\varepsilon)\psi_n(r)$, and write the radial spinor components as

$$\psi_n(r) = \begin{pmatrix} \phi_n^+(r) \\ \phi_n^-(r) \end{pmatrix} \tag{2.12}$$

Contrary to the basis set used in Ref. [11], these basis elements are taken to be energy independent. The upper radial spinor component reads as follows

$$\phi_n^+(r) = \begin{cases} a_n^+ x^{\gamma+1} e^{-x/2} L_n^{\nu_+}(x) &, \kappa > 0 \\ a_n^- x^{-\gamma} e^{-x/2} L_n^{\nu_-}(x) &, \kappa < 0 \end{cases} \tag{2.13}$$

where $x = \omega r$, $L_n^\nu(x)$ are the associated Laguerre polynomials of order $n$ [12], and the normalization constants are $a_n^\pm = \sqrt{\omega\Gamma(n+1)/\Gamma(n+\nu_\pm+1)}$. The basis parameters $\omega$ and $\nu_\pm$ are real such that $\omega > 0$, $\nu_\pm > -1$. We are seeking a two-component spinor basis that supports a tridiagonal matrix representation for the reference Hamiltonian $H_0$. To that end, the lower spinor component should be related to the upper by the "kinetic balance" relation, which is suggested by the reference wave equation $(H_0 - \varepsilon)\psi_n = 0$. That is,

$$\phi_n^-(r) = \tfrac{\lambdabar}{\mu}\left(-\tfrac{Z}{\kappa} + \tfrac{\gamma}{r} + \tfrac{d}{dr}\right)\phi_n^+(r), \tag{2.14}$$

where $\mu$ is another real basis parameter. In Ref. [11], exact solvability requirement of the reference (Dirac-Coulomb) problem resulted in the energy-dependent basis by dictating that $\mu = \varepsilon + \gamma/\kappa$. However, here we are interested only in an approximate solution to the

–5–

full problem that still includes an arbitrary, but short-range, scattering potential $\mathcal{V}(r)$. Therefore, for practical calculations we insist that the parameter $\mu$ be energy-independent. One can show that the spinor basis defined by Eq. (2.13) and Eq. (2.14) results in a tridiagonal matrix representation for the reference Hamiltonian $H_0$ only if $v_\pm = \pm(2\gamma + 1)$ = $2|\gamma| \pm 1$. Moreover, taking the limit $\hbar \to 0$ of the matrix elements of $H_0$ gives the correct nonrelativistic matrix elements of the Coulomb Hamiltonian only if $\mu = 2$ [13]. Using the differential formula and recursion relations of the Laguerre polynomials [12] in the kinetic balance relation (2.14) with $\mu = 2$, we obtain the following

$$\phi_n^-(r) = \frac{\hbar\omega}{4} a_n^\pm x^{\frac{v_\pm - 1}{2}} e^{-x/2} \times$$
$$\left\{ 2\left[\gamma - \frac{Z}{\kappa\omega}(2n + v_\pm + 1)\right] L_n^{v_\pm}(x) - \sigma_-(n + v_\pm) L_{n-1}^{v_\pm}(x) + \sigma_+(n+1) L_{n+1}^{v_\pm}(x) \right\} \quad (2.15)$$

corresponding to $\pm\kappa > 0$ and where $\sigma_\pm = 1 \pm (2Z/\kappa\omega)$. Moreover, using the known relations among Laguerre polynomials of different indices, we can finally rewrite the $L^2$ spinor basis as

$$\psi_n(r) = a_n^+ x^\gamma e^{-x/2} \begin{pmatrix} (n + 2\gamma + 1)L_n^{2\gamma}(x) - (n+1)L_{n+1}^{2\gamma}(x) \\ \frac{\hbar\omega}{4}\left[\sigma_-(n + 2\gamma + 1)L_n^{2\gamma}(x) + \sigma_+(n+1)L_{n+1}^{2\gamma}(x)\right] \end{pmatrix}, \quad \kappa > 0 \quad (2.16a)$$

$$\psi_n(r) = a_n^- x^{-\gamma} e^{-x/2} \begin{pmatrix} L_n^{-2\gamma}(x) - L_{n-1}^{-2\gamma}(x) \\ -\frac{\hbar\omega}{4}\left[\sigma_+ L_n^{-2\gamma}(x) + \sigma_- L_{n-1}^{-2\gamma}(x)\right] \end{pmatrix}, \quad \kappa < 0 \quad (2.16b)$$

Using these and the orthogonality relation of the Laguerre polynomials, we obtain the following tridiagonal basis-overlap matrix (representation of the identity)

$$\langle \psi_n | \psi_m \rangle = \left\{ (2n + v_\pm + 1)\left[1 + (\hbar\omega/4)^2 \rho_+\right] - \hbar^2 \omega Z(\gamma/2\kappa) \right\} \delta_{nm}$$
$$- \left[1 - (\hbar\omega/4)^2 \rho_-\right]\left\{\sqrt{n(n + v_\pm)}\,\delta_{n,m+1} + \sqrt{(n+1)(n + v_\pm + 1)}\,\delta_{n,m-1}\right\} \quad (2.17)$$

corresponding to $\pm\kappa > 0$ and where $\rho_\pm = 1 \pm (2Z/\kappa\omega)^2$. Additionally, the matrix elements of the tridiagonal reference Hamiltonian is calculated for $\pm\kappa > 0$ as

$$(H_0)_{n,m} = \left\{ (2n + v_\pm + 1)\left[\frac{\gamma}{\kappa} + (\hbar\omega/4)^2\left(4 - \frac{\gamma}{\kappa}\right)\rho_+\right] + 2\hbar^2 \omega Z(1 - \gamma/2\kappa)^2 \right\} \delta_{nm}$$
$$- \left[\frac{\gamma}{\kappa} - (\hbar\omega/4)^2\left(4 - \frac{\gamma}{\kappa}\right)\rho_-\right]\left\{\sqrt{n(n + v_\pm)}\,\delta_{n,m+1} + \sqrt{(n+1)(n + v_\pm + 1)}\,\delta_{n,m-1}\right\} \quad (2.18)$$

Now, since the scattering potential $U(r)$ is short-range, then we can assume that it will be well-represented by its matrix elements in a finite $N$-dimensional subset of the basis, $\{\psi_n\}_{n=0}^{N-1}$, for some large enough integer $N$. Therefore, the matrix elements of the potential $U(r)$, which is still arbitrary, can only be evaluated numerically. If we define the $(n,m)$ sampling element of a real square integrable (but not necessarily differentiable) radial function $F(r)$ by the Laguerre polynomials as the value of the integral

$$F_{nm}^\nu = \sqrt{\frac{\Gamma(n+1)\Gamma(m+1)}{\Gamma(n+\nu+1)\Gamma(m+\nu+1)}} \int_0^\infty x^\nu e^{-x} L_n^\nu(x) F(x/\omega) L_m^\nu(x)\, dx, \quad (2.19)$$

Then, we obtain the following potential matrix elements, for $\pm\kappa > 0$

$$\langle \phi_n^+ | U_+ | \phi_m^+ \rangle = R_{nm}^{v_\pm} \quad (2.20a)$$

–6–

$$\left\langle \phi_n^+ \left| U_0 \right| \phi_m^- \right\rangle + \left\langle \phi_n^- \left| U_0 \right| \phi_m^+ \right\rangle =$$

$$\pm \frac{\lambda \omega}{2} \left\{ \sigma_\mp \sqrt{(n_\pm + 2|\gamma|)(m_\pm + 2|\gamma|)} (U_0)_{nm}^{2|\gamma|} - \sigma_\pm \sqrt{n_\pm m_\pm} (U_0)_{n\pm 1, m\pm 1}^{2|\gamma|} \right. \quad (2.20b)$$

$$\left. + \frac{2Z}{|\kappa|\omega} \left[ \sqrt{n_\pm (m_\pm + 2|\gamma|)} (U_0)_{m,n\pm 1}^{2|\gamma|} + \sqrt{m_\pm (n_\pm + 2|\gamma|)} (U_0)_{n,m\pm 1}^{2|\gamma|} \right] \right\}$$

$$\left\langle \phi_n^- \left| U_- \right| \phi_m^- \right\rangle =$$

$$\left( \frac{\lambda \omega}{4} \right)^2 \left\{ \sigma_\mp^2 \sqrt{(n_\pm + 2|\gamma|)(m_\pm + 2|\gamma|)} (U_-)_{nm}^{2|\gamma|} + \sigma_\pm^2 \sqrt{n_\pm m_\pm} (U_-)_{n\pm 1, m\pm 1}^{2|\gamma|} \right. \quad (2.20c)$$

$$\left. + \rho_- \left[ \sqrt{n_\pm (m_\pm + 2|\gamma|)} (U_-)_{m,n\pm 1}^{2|\gamma|} + \sqrt{m_\pm (n_\pm + 2|\gamma|)} (U_-)_{n,m\pm 1}^{2|\gamma|} \right] \right\}$$

where the radial function $R(r)$ in Eq. (2.20a) is defined as $R(r) = (\omega r) U_+(r)$ and $n_\pm = n + \frac{1 \pm 1}{2}$. Now, for an integer $K$ larger than the chosen size of the basis $N$, we can use Gauss quadrature [9] to give the following approximate evaluation for the integral (2.19)

$$F_{nm}^\nu \cong \sum_{k=0}^{K-1} \Lambda_{nk}^\nu \Lambda_{mk}^\nu F(\eta_k^\nu / \omega), \quad (2.21)$$

where $\{\Lambda_{nk}^\nu\}_{n=0}^{K-1}$ is the normalized eigenvector associated with the eigenvalue $\eta_k^\nu$ of the $K \times K$ tridiagonal symmetric matrix of the quadrature associated with the Laguerre polynomials $\{L_n^\nu\}$. That is, the matrix whose $(n,m)$ elements are $(2n + \nu + 1)\delta_{nm} + \sqrt{n(n+\nu)}\, \delta_{n,m+1} + \sqrt{(n+1)(n+\nu+1)}\, \delta_{n,m-1}$ for $n, m = 0, 1, 2, ..., K-1$.

In the following section, we show that the representation obtained above is compatible and, in fact, very useful in obtaining the relativistic bound states and resonance energies using the complex scaling method.

### III. RELATIVISTIC BOUND STATES AND RESONANCE ENERGIES

Bound states for the one-particle relativistic problem described by the Hamiltonian (2.1) are for energies less than the rest mass. That is the bound states energy spectrum is confined to the real energy interval $\varepsilon \in [-1, +1]$. Moreover, the time dependence of the total spinor wavefunction for positive (negative) energy, which is given by the exponential factor $e^{-i\varepsilon t}$ ($e^{+i\varepsilon t}$), implies that the imaginary part of resonance energies associated with hermitian Hamiltonians should be negative (positive). Therefore, resonance energies will be located in the second and fourth quarter of the complex energy plane with $|\text{Re}(\varepsilon)| > 1$. Consequently, there will be two energy thresholds for relativistic scattering. These are $\varepsilon = +1$ and $\varepsilon = -1$ for positive and negative energy, respectively. Moreover, there will be two semi-infinite cut lines on the real $\varepsilon$-axis corresponding to the continuum scattering states. One for $\varepsilon > +1$ and the other for $\varepsilon < -1$. Thus, the relativistic problem resembles a two-channel nonrelativistic problem (but one with negative energy). Complex scaling causes the two cut lines to rotate clockwise around the origin ($\varepsilon = 0$) with an angle $\theta$ [5,6,8]. Therefore, these lines will be deformed curving up (down) for energies near $\varepsilon = -1$ ($\varepsilon = +1$). If we parameterize the two cut lines by a real parameter $\xi$ then the continuous rotated spectrum could be written as the following set of complex numbers [6,8]:

$$z(\xi) = \pm\sqrt{\xi^2 e^{-2i\theta} + 1}. \tag{3.1}$$

For energies close to $\varepsilon = \pm 1$ (i.e., in the nonrelativistic limit where $\xi$ is small), these complex scaled curves slope by the angle $2\theta$ as seen from Eq. (3.1). Figure 3 shows the effect of the complex scaling transformation $r \to re^{i\theta}$ on the spectrum of $H$. Bound states are shown as blue dots on the real energy axis whereas scattering states are designated by the red dots. The two rotated cut "lines" are the black solid curves. We choose to write all energy values that will be given below in terms of the relativistic energy variable

$$\mathcal{E}(\varepsilon) = (\varepsilon^2 - 1)/2\lambdabar^2. \tag{3.2}$$

Therefore, the continuous energy spectrum (the cut lines of the relativistic Green's function) in terms of this complex variable is simply written as $\mathcal{E}(\xi) = \frac{1}{2}(\xi/\lambdabar)^2 e^{-2i\theta}$. This could also be written as $\mathcal{E} = \frac{1}{2}k^2$, where $k = (\xi/\lambdabar)e^{-i\theta}$ is the complex "relativistic wave number". In the nonrelativistic limit ($\lambdabar \to 0, \varepsilon \to 1 + \lambdabar^2 E$) $\mathcal{E}$ will, in fact, be equal to the nonrelativistic energy $E$.

Now, we show that the matrix representation of the Hamiltonian (2.1) in the spinor basis obtained in the previous section results in a simple and straightforward scheme for the implementation of the complex scaling method on the relativistic problem. To that end, we write the matrix elements of the total Hamiltonian as

$$\begin{aligned}
H_{nm} = \langle \psi_n(x) | H | \psi_m(x) \rangle &= \langle \phi_n^+(x) | \tfrac{\gamma}{\kappa} + 2\lambdabar^2 \omega \tfrac{Z}{x} | \phi_m^+(x) \rangle - \tfrac{\gamma}{\kappa} \langle \phi_n^-(x) | \phi_m^-(x) \rangle \\
&+ \lambdabar \omega \left[ \langle \phi_n^+(x) | -\tfrac{Z}{\kappa\omega} + \tfrac{\gamma}{x} - \tfrac{d}{dx} | \phi_m^-(x) \rangle + \langle \phi_n^-(x) | -\tfrac{Z}{\kappa\omega} + \tfrac{\gamma}{x} + \tfrac{d}{dx} | \phi_m^+(x) \rangle \right] \\
&+ \lambdabar^2 \langle \phi_n^+(x) | U_+(x/\omega) | \phi_m^+(x) \rangle + \lambdabar^2 \langle \phi_n^-(x) | U_-(x/\omega) | \phi_m^-(x) \rangle \\
&+ \lambdabar \left[ \langle \phi_n^+(x) | U_0(x/\omega) | \phi_m^-(x) \rangle + \langle \phi_n^-(x) | U_0(x/\omega) | \phi_m^+(x) \rangle \right]
\end{aligned} \tag{3.3}$$

where $x = \omega r$. It is, thus, obvious that the effect of the transformation $r \to re^{i\theta}$ on the differential matrix operator $H(r)$ in configuration space is equivalent to the transformation $\omega \to \omega e^{-i\theta}$ of its matrix elements in the spinor representation (2.16). Now, this latter transformation is very easy to implement on the matrix elements (2.17), (2.18) and (2.20) that make up the representation (3.3). One may also note that the $\sqrt{\omega}$ factor in the normalization constant $a_n^\pm$ together with the integration measure $dr$ cooperate to preserve the equivalence between these two transformations by producing the integration measure $dx$.

We start with a consistency check of the results obtained using the representation given in the previous section against known exact results. For that purpose we calculate the relativistic energy of the bound states for Hydrogen-like ions (where $\lambdabar = \alpha a_0$ and $\mathcal{V} = 0$) and compare that with the exact formula [10]

$$\varepsilon_n = \pm \left[ 1 + \left( \frac{\lambdabar Z}{n + \tau + 1} \right)^2 \right]^{-1/2}, \tag{3.4}$$

where $\tau = \begin{cases} \gamma & ,\kappa > 0 \\ -\gamma - 1 & ,\kappa < 0 \end{cases}$. Throughout our calculation, we use the following strategy. We search for a range of values of the basis scale parameter $\omega$ that give stable results for a given basis size $N$. We calculate the average result corresponding to several values of $\omega$ (preferably, from the middle of the range) keeping only significant digits. Those are the digits that do not change by changing $N$ by, say, 5%. Table I shows a good agreement relative to a basis size $N = 200$ and with calculation stability for a range of values of $\omega = 1$ to 5 a.u. Nonetheless, built on a general-purpose software package and by an author with limited programming experience, the code used in the calculation is not meant to achieve high precision but to demonstrate the utility and applicability of the extended method. Of course, very high precision could be achieved with better computational routines to find, for example, the generalized eigenvalues and eigenvectors for large tridiagonal matrices. Moreover, since the relativistic energy spectrum (for typical systems) clusters around $\varepsilon = \pm 1$, then to reduce computational errors it is advisable in some cases to calculate the eigenvalues $\varepsilon_n \mp 1$ instead of $\varepsilon_n$, respectively. This is also one of the reasons that we chose to give our results in terms of the energy variable $\mathcal{E}$ instead of $\varepsilon$.

To illustrate further the utility and accuracy of the method, we implement it on an atomic model with known nonrelativistic resonance structure. We choose the well-established and frequently used model $V(r) = 7.5\, r^2 e^{-r}$ [14-19] and take $S(r) = W(r) = 0$ in the potential matrix (2.5). Figure 4 is a snap shot from a video that shows how the positive energy resonances become exposed as the cut "line" sweeps the lower half of the complex $\mathcal{E}$-plane, while the angle $\theta$ of the complex scaling transformation gradually increases from 0.0 to 0.9 radians. The snap shot is taken at $\theta = 0.7$ radians. It should be clear that, just like in nonrelativistic scattering, the cut line gets deformed near resonances due to the localization effect. In Table II we compare the nonrelativistic limit of our calculation of the resonance energies associated with this model potential for $\kappa = -1$ ($\ell = 0$) against known nonrelativistic results elsewhere. In the calculation, the nonrelativistic limit was achieved by taking $\hbar/a_0 = \alpha/100$. That is, the fine structure constant was effectively reduced by a factor of 100 or, equivalently, the speed of light was ascribed a value 100 times larger. We took a basis size $N = 100$ and calculation stability is for a range of values of $\omega = 5$ to 30 a.u. Here again, we note the good agreement despite the relatively small size of basis (compared to other studies where five to ten fold larger sizes are typical) and limited programming power. It is worth noting that computational errors increase substantially if one attempts to obtain the nonrelativistic limit by reducing the value of $\hbar$ too much (for example, by taking $\hbar/a_0 = 10^{-4}\alpha$). Table III gives a more comprehensive list of the relativistic energy resonances ($\mathcal{E} = \mathcal{E}_r - i\Lambda_r/2$) for the same potential as in Table II, but for several values of $Z$ and $\kappa$.

As an additional advantage of the tridiagonal representation constructed in the previous section, we use it in obtaining highly accurate values of the relativistic bound states energies for the Woods-Saxon potential as a model of nuclear interaction. We consider a system consisting of a nucleon and a heavy nucleus of mass number $A \gg 1$ and atomic number $Z$. In addition to the Coulomb interaction between the nucleon (proton) and the nucleus, we model the nuclear interaction by the Woods-Saxon potential

$$V_{\text{WS}}(r) = \frac{-V_0}{1 + e^{(r-R_0)/r_0}}, \qquad (3.5)$$

for a given set of parameters $V_0$, $R_0$, and $r_0$, where $r_0 \ll R_0 \propto A^{1/3}$ [20]. Additionally, for proton scattering, we model the nucleus as a sphere of a uniform charge distribution $Ze$ with charge radius $R_c$. That is, we write the electrostatic potential as

$$V_C(r) = \lambda^2 \begin{cases} \dfrac{Z}{r} & , r \geq R_c \\ \dfrac{Z}{2R_c}\left[3 - \left(\dfrac{r}{R_c}\right)^2\right] & , r < R_c \end{cases} \qquad (3.6)$$

Now, there are three different possibilities for the Woods-Saxon potential coupling and combinations thereof. These are the vector, scalar, and pseudo-scalar couplings as shown in Eq. (2.5). That is, for a given choice of dimensionless coupling parameters $\{\eta_i\}_{i=V,S,W}$, we take $V(r) = \eta_V V_{\text{WS}}(r)$, $S(r) = \eta_S V_{\text{WS}}(r)$, and $W(r) = \eta_W V_{\text{WS}}(r)$. However, in the non-relativistic limit the vector and scalar coupling produce the same results, which could also be verified numerically. We start by calculating the bound states energy spectrum for pure vector coupling and for a given choice of physical parameters. We also compare the non-relativistic limit of our calculation to the exact spectrum, which is known only for $Z = 0$ (e.g., for the neutron) and $\kappa = -1$ ($\ell = 0$) [21]. Table IV lists these results for $N = 100$ showing a good agreement (to seven significant digits) with calculation stability relative to variations in $\omega$ in the range 3 to 30 fm$^{-1}$. The nonrelativistic limit was achieved by reducing the value of $\lambda$ 1000 times. Table V gives a more comprehensive list of the relativistic bound state energies for the same potential parameters with $Z = 0$ but for different values of $\kappa$ and for the three types of coupling. In the pseudo-scalar coupling case, it is interesting to observe that: (1) for $\kappa < 0$, there is only one bound state for this configuration, and (2) for $\kappa = +2$, one of the bound state energies is greater then $V_0$. The last observation could be understood by noting that $W(r)$ belongs to the odd part of the Hamiltonian matrix whereas the energy $\varepsilon$ belongs to the even part, consequently its contribution to the energy is of the form $W^2$ and/or $\pm W'$ which could exceed $V_0$. This is similar to the known properties of potentials in supersymmetric quantum mechanics [22]. Table VI shows similar results to that in Table V but for the proton where the Coulomb interaction (3.6) comes into play in addition to the Woods-Saxon potential. We took $R_c = 1.2 R_0$ and $Z = 50$.

## IV. DISCUSSION

In our calculation of bound states and resonance energies, we used a finite $N$-dimensional basis. However, the matrix representation of the reference Hamiltonian $\mathcal{H}_0$ is fully accounted for as given by Eq. (2.18). It is only the potential matrix $\mathcal{V}$ that has to be approximated by its elements in the finite subset of the basis as given by Eqs. (2.20a-c). Therefore, the matrix representation of the Hamiltonian, which is available at our disposal, is the following

$$\mathcal{H}_{nm} \cong \begin{cases} (\mathcal{H}_0)_{nm} + \mathcal{V}_{nm} & , n,m \leq N-1 \\ (\mathcal{H}_0)_{nm} & , n,m > N-1 \end{cases} \qquad (4.1)$$

Consequently, if one could find the means of handling the infinite tridiagonal tail of this matrix and thus account for the full contribution of $\mathcal{H}_0$ in the calculation, then one should expect to obtain much more accurate results. In fact, such a scheme does exist. The representation (4.1) is the fundamental underlying structure of the J-matrix method [23]. It is an algebraic method of quantum scattering. The method takes into account the full contribution of the reference Hamiltonian analytically. It is in our work plan to combine the relativistic extension of the complex scaling method developed here and the relativistic version of the J-matrix method [13,24] for achieving a substantial improvement on the accuracy for calculating the relativistic energy of bound states and resonances.

## ACKNOWLEDGMENT

I am grateful to M. S. Abdelmonem for providing some of the reference material cited in the Tables. The support provided by the KFUPM library in literature search is highly appreciated. This work is partially supported by KFUPM grant FT-2006/06.

**TABLE CAPTIONS:**

**Table I**: A comparison of the relativistic energy of bound states for Hydrogen-like ions obtained using the tridiagonal representation in Eq (2.18) against the exact formula (3.4). The energy variable $\mathcal{E}$ is related to $\varepsilon$ by Eq. (3.2) and is measured in units of $mc^2$. The basis size $N = 200$.

**Table II**: The nonrelativistic limit of our calculation of the resonance energies, $E_r - \frac{i}{2}\Gamma_r$, associated with the model potential $V(r) = 7.5r^2 e^{-r}$, $S(r) = W(r) = 0$ and for $\kappa = -1$ ($\ell = 0$) against known nonrelativistic results elsewhere. We took a basis size $N = 100$ and the nonrelativistic limit was achieved by scaling down the fine structure constant 100 times.

**Table III**: A more comprehensive list of the relativistic energy resonances $\mathcal{E}_r - i\Lambda_r/2$ for the same potential as in Table II, but for several values of $Z$ and $\kappa$.

**Table IV**: The relativistic S-wave bound state energies (2nd column) of a neutron in a heavy nucleus where the nuclear interaction is modeled by the Woods-Saxon potential in a vector coupling only and for $V_0 = 300$ MeV, $R_0 = 7.0$ fm, and $r_0 = 0.5$ fm. The non-relativistic limit (3rd column) is compared with the exact values (4th column). We set the basis size $N = 100$.

**Table V**: A more comprehensive list of the relativistic bound state energies ($-\mathcal{E}$, in MeV) for the same potential parameters as in Table IV with $Z = 0$ (the neutron) but for different values of angular momentum and for the three types of coupling.

**Table VI**: Similar to the results in Table V but for the proton where the Coulomb interaction, modeled by the uniform charge sphere in Eq. (3.6), comes into play in addition to the Woods-Saxon potential. We took $V_0 = 300$ MeV, $R_0 = 7.0$ fm, $r_0 = 0.5$ fm, $Z = 50$, and $R_c = 1.2 R_0$.

**FIGURE CAPTIONS:**

**Fig. 1** (Color online): The full effect of the complex scaling (rotation) $r \to re^{i\theta}$ on the bound states (blue dots), continuum states (black line), and resonances (red dots) in the nonrelativistic complex energy plane.

**Fig. 2**: Showing the same as Fig. 1 but by means of a finite dimensional basis. The cut line is replaced by a string of dots (eigenvalues of the finite complex rotated Hamiltonian) which is slightly deformed in the neighborhood of resonances.

**Fig. 3** (Color online): The full effect of the transformation $r \to re^{i\theta}$ on the spectrum of the complex rotated Dirac Hamiltonian. Bound states are the blue dots on the real energy $\varepsilon$-axis whereas scattering states are the red dots. The rotated positive and negative energy continuum are the two black solid curves.

**Fig. 4** (958 KB MPG): A snap shot from a video revealing some of the resonance poles in the complex $\mathcal{E}$-plane associated with the potential $V(r) = 7.5r^2 e^{-r}$, $S(r) = W(r) = 0$ and for $Z = 1$, $\kappa = -1$. The video shows how the positive energy resonances become exposed as the cut "line" sweeps the lower half of the complex $\mathcal{E}$-plane, while the angle $\theta$ of the complex scaling transformation gradually increases from 0.0 to 0.9 radians. The snap shot is taken at $\theta = 0.7$ radians and the basis size $N = 100$.

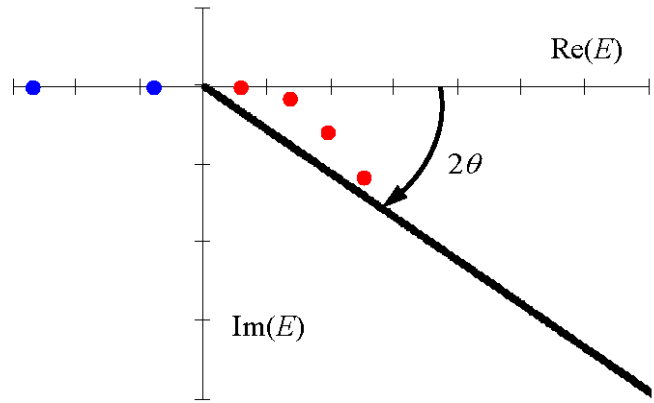

**Fig. 1**

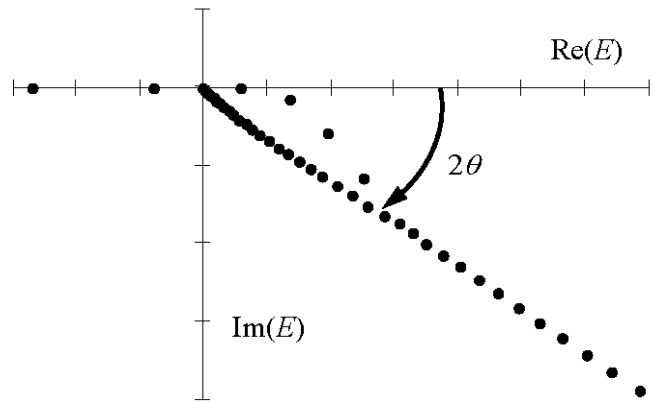

**Fig. 2**

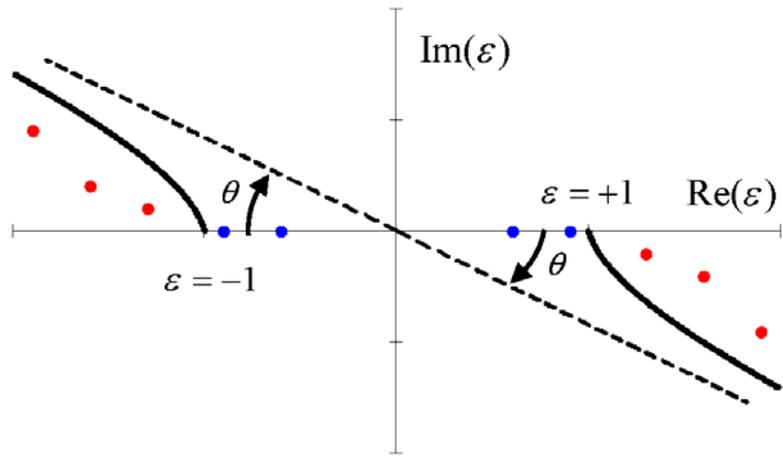

**Fig.3**

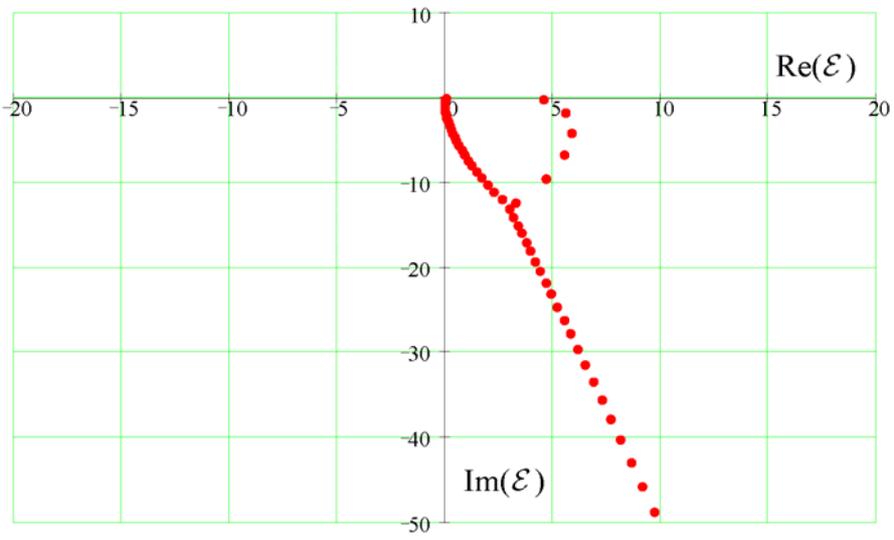

**Fig. 4**

**Table I**

| Z | j | κ | State | −ℰ (exact) | −ℰ (RCS-07.1) |
|---|---|---|---|---|---|
| −1 | | | | | |
| | $\frac{1}{2}$ | −1 | $1S_{1/2}$ | 0.500 000 000 000 | 0.500 000 001 42 |
| | | | $2S_{1/2}$ | 0.125 001 664 149 | 0.125 001 664 38 |
| | | | $3S_{1/2}$ | 0.055 556 212 996 | 0.055 556 213 18 |
| | | +1 | $2P_{1/2}$ | 0.125 001 664 149 | 0.125 001 664 39 |
| | | | $3P_{1/2}$ | 0.055 556 212 996 | 0.055 556 213 14 |
| | | | $4P_{1/2}$ | 0.031 250 312 026 | 0.031 250 312 15 |
| | $\frac{3}{2}$ | −2 | $2P_{3/2}$ | 0.125 000 000 000 | 0.125 000 000 02 |
| | | | $3P_{3/2}$ | 0.055 555 719 912 | 0.055 555 719 93 |
| | | | $4P_{3/2}$ | 0.031 250 104 007 | 0.031 250 104 00 |
| | | +2 | $3D_{3/2}$ | 0.055 555 719 912 | 0.055 555 719 92 |
| | | | $4D_{3/2}$ | 0.031 250 104 007 | 0.031 250 104 01 |
| | | | $5D_{3/2}$ | 0.020 000 063 902 | 0.020 000 063 90 |
| −2 | | | | | |
| | $\frac{1}{2}$ | −1 | $1S_{1/2}$ | 2.000 000 000 000 | 2.000 000 090 73 |
| | | | $2S_{1/2}$ | 0.500 026 628 513 | 0.500 026 644 03 |
| | | | $3S_{1/2}$ | 0.222 232 742 066 | 0.222 232 751 40 |
| | | +1 | $2P_{1/2}$ | 0.500 026 628 513 | 0.500 026 644 02 |
| | | | $3P_{1/2}$ | 0.222 232 742 066 | 0.222 232 751 40 |
| | | | $4P_{1/2}$ | 0.125 004 992 779 | 0.125 005 000 38 |
| | $\frac{3}{2}$ | −2 | $2P_{3/2}$ | 0.500 000 000 000 | 0.500 000 001 41 |
| | | | $3P_{3/2}$ | 0.222 224 851 984 | 0.222 224 852 41 |
| | | | $4P_{3/2}$ | 0.125 001 664 149 | 0.125 001 664 39 |
| | | +2 | $3D_{3/2}$ | 0.222 224 851 984 | 0.222 224 852 42 |
| | | | $4D_{3/2}$ | 0.125 001 664 149 | 0.125 001 664 39 |
| | | | $5D_{3/2}$ | 0.080 001 022 452 | 0.080 001 022 63 |

**Table II**

| $Z$ | $E_r$ (a.u.) | $\Gamma_r$ (a.u.) | Reference |
|---|---|---|---|
| 0 | 3.426 39 | 0.0255 49 | [14] |
|   | 3.425 7 | 0.0256 | [15] |
|   | 3.426 | 0.0256 | [16] |
|   | 3.426 | 0.0258 | [17] |
|   | 3.426 390 331 | 0.0255 489 62 | [18] |
|   | 3.426 390 310 | 0.0255 489 61 | [19] |
|   | 3.426 390 3 | 0.0255 49 | this work |
| 0 | 4.834 806 841 | 2.235 753 338 | [18] |
|   | 4.834 806 841 | 2.235 753 338 | [19] |
|   | 4.834 806 9 | 2.235 752 9 | this work |
| 0 | 5.277 279 780 | 6.778 106 356 | [18] |
|   | 5.277 279 864 | 6.778 106 591 | [19] |
|   | 5.277 279 8 | 6.778 106 5 | this work |
| −1 | 1.780 5 | $9.58 \times 10^{-5}$ | [17] |
|   | 1.780 524 536 | $9.571\,9 \times 10^{-5}$ | [18] |
|   | 1.780 524 536 | $9.571\,94 \times 10^{-5}$ | [19] |
|   | 1.780 525 | $9.59 \times 10^{-5}$ | this work |
| −1 | 4.101 494 946 | 1.157 254 428 | [18] |
|   | 4.101 494 946 | 1.157 254 428 | [19] |
|   | 4.101 495 | 1.157 254 | this work |
| −1 | 4.663 461 099 | 5.366 401 539 | [18] |
|   | 4.663 461 097 | 5.366 401 540 | [19] |
|   | 4.663 461 | 5.366 402 | this work |

**Table III**

| | $\kappa = -1\ (j=\tfrac{1}{2}, \ell=0)$ | | $\kappa = +1\ (j=\tfrac{1}{2}, \ell=1)$ | | $\kappa = -2\ (j=\tfrac{3}{2}, \ell=1)$ | |
|---|---|---|---|---|---|---|
| Z | $\mathcal{E}_r$ (a.u.) | $\Lambda_r$ (a.u.) | $\mathcal{E}_r$ (a.u.) | $\Lambda_r$ (a.u.) | $\mathcal{E}_r$ (a.u.) | $\Lambda_r$ (a.u.) |
| 0 | 2.9465842 | 23.06811 | 2.2170 | 25.8544 | 2.2170 | 25.8545 |
|  | 3.4266874221 | 0.0255518009 | 3.80129781 | 20.19320406 | 3.801256006 | 20.19329161 |
|  | 4.2687950416 | 17.439000786 | 4.647141064 | 0.6506112695 | 4.6471912798 | 0.6506993408 |
|  | 4.8354225415 | 2.2361196639 | 4.888143904 | 14.627813147 | 4.88811024012 | 14.627927250 |
|  | 5.0654945401 | 11.955326382 | 5.361234056 | 4.3953039551 | 5.36124186636 | 4.3954504642 |
|  | 5.2780344286 | 6.7796704926 | 5.428512321 | 9.283011964 | 5.42849401850 | 9.2831485326 |
| −1 | 1.7805455986 | 0.0000957022 | 1.92025 | 24.5794 | 1.92010 | 24.5795 |
|  | 2.592623 | 21.393386 | 3.453842379 | 18.973863298 | 3.4537211666 | 18.974083056 |
|  | 3.8493853286 | 15.820609152 | 3.8482370611 | 0.2752972793 | 3.8483561762 | 0.2753948771 |
|  | 4.1018900840 | 1.1572503194 | 4.4767397212 | 13.480474062 | 4.47665785037 | 13.480735107 |
|  | 4.5617673176 | 10.415641527 | 4.7506422650 | 3.5058292798 | 4.75066932711 | 3.5061036576 |
|  | 4.6640968354 | 5.3671942790 | 4.9336803340 | 8.2353324063 | 4.93364679163 | 8.2356163021 |
| +1 | 3.281076 | 24.715461 | 2.503217 | 27.12133 | 2.503265 | 27.12135 |
|  | 4.5950476252 | 0.25794882226 | 4.1363561854 | 21.403666589 | 4.136385087 | 21.403689968 |
|  | 4.6664918654 | 19.0235931997 | 5.2842961861 | 15.766367118 | 5.2843094425 | 15.766398357 |
|  | 5.54553715604 | 13.4562601372 | 5.3726029555 | 1.1464590251 | 5.3726223834 | 1.1465226325 |
|  | 5.57049606384 | 3.42295459433 | 5.9032040396 | 10.323528755 | 5.9032078368 | 10.323575489 |
|  | 5.86827722644 | 8.1611145838 | 5.9418167716 | 5.2816442304 | 5.9418208860 | 5.2817101553 |

**Table IV**

| n | $-\mathcal{E}$ (MeV) | $-\mathcal{E}\ (\lambdabar \to 0)$ | $-E$ (exact) |
|---|---|---|---|
| 0 | 248.097 145 52 | 294.140 89 | 294.140 931 652 |
| 1 | 237.054 887 98 | 278.114 19 | 278.114 215 979 |
| 2 | 220.196 163 75 | 253.947 64 | 253.947 661 400 |
| 3 | 198.227 425 13 | 222.949 65 | 222.949 660 882 |
| 4 | 171.696 790 31 | 186.185 39 | 186.185 386 283 |
| 5 | 141.193 394 59 | 144.774 73 | 144.774 714 174 |
| 6 | 107.473 683 13 | 100.156 91 | 100.156 894 346 |
| 7 | 71.636 866 80 | 54.636 39 | 54.636 364 679 |
| 8 | 35.601 083 66 | 13.500 14 | 13.500 142 641 |
| 9 | 4.654 4 | ......... | ......... |

**Table V**

|  | $\kappa = -1$ | $\kappa = +1$ | $\kappa = -2$ | $\kappa = +2$ |
|---|---|---|---|---|
| vector coupling | 248.097 145 52 | 244.076 127 32 | 244.059 446 71 | 239.119 541 66 |
|  | 237.054 887 98 | 229.938 096 34 | 229.899 015 05 | 222.014 489 51 |
|  | 220.196 163 75 | 210.410 120 18 | 210.344 267 31 | 199.900 429 87 |
|  | 198.227 425 13 | 186.063 697 07 | 185.968 185 38 | 173.254 907 27 |
|  | 171.696 790 31 | 157.439 926 21 | 157.313 672 75 | 142.637 914 63 |
|  | 141.193 394 59 | 125.187 949 62 | 125.032 274 33 | 108.783 513 19 |
|  | 107.473 683 13 | 90.195 744 78 | 90.015 480 66 | 72.759 805 80 |
|  | 71.636 866 80 | 53.852 584 38 | 53.658 868 15 | 36.409 597 85 |
|  | 35.601 083 66 | 18.998 020 08 | 18.820 187 69 | 4.636 7 |
|  | 4.654 4 | ………. | ………. | ………. |
| scalar coupling | 247.088 024 55 | 242.009 800 35 | 242.030 318 33 | 235.733 821 71 |
|  | 233.170 516 14 | 224.097 631 24 | 224.146 241 12 | 213.985 976 67 |
|  | 211.784 368 37 | 199.221 854 88 | 199.304 301 40 | 185.745 633 06 |
|  | 183.796 140 07 | 168.127 814 75 | 168.247 719 57 | 151.699 962 08 |
|  | 149.983 193 01 | 131.672 656 04 | 131.830 529 80 | 112.846 253 19 |
|  | 111.385 150 70 | 91.148 100 50 | 91.339 148 21 | 70.834 830 40 |
|  | 69.699 134 47 | 48.855 361 44 | 49.063 450 08 | 28.950 607 86 |
|  | 28.330 179 49 | 10.382 286 3 | 10.549 754 | ………. |
| pseudo-scalar coupling | 24.969 391 16 | 209.826 279 93 | 11.553 984 | 313.824 097 95 |
|  |  | 106.891 049 87 |  | 209.827 142 54 |
|  |  | 69.112 179 61 |  | 146.523 762 21 |
|  |  | 61.629 018 89 |  | 109.198 608 21 |
|  |  | 36.898 898 62 |  | 93.037 202 30 |
|  |  | 13.710 545 4 |  | 67.109 171 25 |
|  |  | ………. |  | 36.347 124 47 |
|  |  | ………. |  | 6.957 939 |

**Table VI**

|  | $\kappa = -1$ | $\kappa = +1$ | $\kappa = -2$ | $\kappa = +2$ |
|---|---|---|---|---|
| vector coupling | 239.563 803 170 | 235.622 595 549 | 235.607 266 173 | 230.691 823 560 |
|  | 228.427 148 185 | 221.250 580 826 | 221.212 660 865 | 213.257 784 117 |
|  | 211.338 683 118 | 201.429 688 194 | 201.364 600 016 | 190.794 499 205 |
|  | 189.054 177 508 | 176.728 212 555 | 176.633 055 235 | 163.758 365 131 |
|  | 162.150 372 138 | 147.708 047 152 | 147.581 800 246 | 132.725 735 38 |
|  | 131.243 257 14 | 115.044 987 76 | 114.889 116 00 | 98.458 924 2 |
|  | 97.121 644 5 | 79.665 374 0 | 79.485 004 6 | 62.075 202 |
|  | 60.937 643 | 43.036 007 | 42.843 025 | 25.545 91 |
|  | 24.749 35 | 8.296 41 | 8.124 16 | ......... |
| scalar coupling | 238.556 334 402 | 233.544 307 761 | 233.569 264 591 | 227.275 002 469 |
|  | 224.495 382 012 | 215.329 897 357 | 215.383 392 390 | 205.114 952 428 |
|  | 202.804 046 574 | 190.075 318 001 | 190.163 351 087 | 176.431 455 638 |
|  | 174.410 677 933 | 158.532 720 69 | 158.658 937 74 | 141.900 936 21 |
|  | 140.134 495 60 | 121.596 415 34 | 121.761 209 39 | 102.558 486 93 |
|  | 101.064 652 4 | 80.617 438 6 | 80.815 653 7 | 60.132 805 |
|  | 58.983 495 | 38.029 052 | 38.243 362 | 18.147 74 |
|  | 17.555 33 | 0.108 88 | 0.270 41 | ......... |
| pseudo-scalar coupling | 15.665 50 | 199.916 335 725 | 2.529 0 | 305.924 565 954 |
|  |  | 95.557 943 017 |  | 200.016 684 061 |
|  |  | 59.234 063 046 |  | 135.940 910 601 |
|  |  | 50.510 465 1 |  | 98.783 396 87 |
|  |  | 25.936 922 |  | 83.036 798 9 |
|  |  | 3.090 3 |  | 56.444 298 |
|  |  | ......... |  | 25.533 513 |